\documentclass[aps, prd, twocolumn, superscriptaddress, nofootinbib, amsmath, amssymb]{revtex4}
\usepackage{graphicx}  
\usepackage{mathrsfs} 
\usepackage[normalem]{ulem} 
\usepackage{amsfonts}
\newcommand{\Tr}{\ensuremath{\mathrm{Tr}}}

\begin{document}

\title{$\theta$-dependence and center symmetry in Yang-Mills theories} 

\author{Claudio Bonati}
\email{claudio.bonati@unipi.it}
\affiliation{Dipartimento di Fisica dell'Universit\`a di Pisa and INFN
  - Sezione di Pisa, Largo Pontecorvo 3, I-56127 Pisa, Italy.}

\author{Marco Cardinali}
\email{marco.cardinali@pi.infn.it}
\affiliation{Dipartimento di Fisica dell'Universit\`a di Pisa and INFN
  - Sezione di Pisa, Largo Pontecorvo 3, I-56127 Pisa, Italy.}

\author{Massimo D'Elia}
\email{massimo.delia@unipi.it}
\affiliation{Dipartimento di Fisica dell'Universit\`a di Pisa and INFN
  - Sezione di Pisa, Largo Pontecorvo 3, I-56127 Pisa, Italy.}

\author{Fabrizio Mazziotti}
\email{fabrizio.mazziotti@pi.infn.it}
\affiliation{Dipartimento di Fisica dell'Universit\`a di Pisa and INFN
  - Sezione di Pisa, Largo Pontecorvo 3, I-56127 Pisa, Italy.}

\begin{abstract}
We investigate the relation between the realization of center symmetry
and the dependence on the topological parameter $\theta$ in $SU(N)$
Yang-Mills theories, exploiting trace deformations as a tool to regulate
center symmetry breaking in a theory with a small compactified direction.
We consider, in particular, $SU(4)$ gauge theory, which admits two possible 
independent deformations, and study, as a first
step, its phase diagram in the deformation plane for two values
of the inverse compactified radius going up to $L^{-1} \sim 500$~MeV, 
comparing the predictions of the effective 1-loop potential of the Polyakov loop with
lattice results.
The $\theta$-dependence of the various phases is then addressed, up to 
the fourth order in $\theta$, by numerical
simulations: results are found to coincide, within
statistical errors, with those of the standard confined phase iff
center symmetry is completely restored and independently of the particular
way this happens, i.e.~either by local suppression of the Polyakov loop
traces or by long range disorder.
\end{abstract}

\maketitle

\section{Introduction} 
\label{intro}

Pure gauge theories, defined on a space-time with one or more compactified
direction, possess a symmetry under global transformations which can be
classified as gauge transformations respecting the periodicity but for a global
element of the center of the gauge group (e.g., $\mathbb{Z}_N$ for $SU(N)$ gauge
theories): this is known as center symmetry.  Such symmetry regulates most of
the phase structure of the pure gauge theory, undergoing spontaneous symmetry
breaking (SSB) for small enough compactification radii, and the Polyakov loop
(holonomy) around the compactified direction is a proper order parameter for
its realization.  When the compactified direction is the thermal Euclidean
direction, the transition is associated to deconfinement and the Polyakov loop
is defined as 
\begin{equation}
P(\vec{x}) = \mathcal{P}\exp\left(i\int_0^{L}A_0(\vec{x}, \tau)
\mathrm{d}\tau\right) \, ;
\label{poldef}
\end{equation}
its trace vanishes in the confined phase ($\langle \mathrm{Tr}P\rangle =0$),
while it is different from zero for $T > T_c$ where $T_c$ is the deconfinement
critical temperature (e.g., for $SU(N)$, $\langle \mathrm{Tr}P\rangle=\alpha
e^{i 2\pi n/N}$, with $n\in\{0, 1, \dots\ N-1\}$ and $\alpha>0$).

Yang-Mills theories are characterized by many other non-perturbative
properties, whose relation to center symmetry is still not clear. Among them, a
significant role is played by the dependence on the topological
parameter $\theta$, which enters the (Euclidean) Lagrangian as follows:
\begin{equation}\label{lagrangian}
\mathcal{L}_\theta  = \frac{1}{4} F_{\mu\nu}^a(x)F_{\mu\nu}^a(x)
- i \theta q(x)\ ,
\end{equation}
where $q(x)$ is the topological charge defined by 
\begin{equation}\label{topchden}
q(x)=\frac{g^2}{64\pi^2} 
\epsilon_{\mu\nu\rho\sigma} F_{\mu\nu}^a(x) F_{\rho\sigma}^a(x)\ .
\end{equation}
A non-zero $\theta$ breaks CP symmetry explicitly, and a non-trivial dependence
on it is induced by gauge configurations with non-trivial winding number $Q =
\int d^4\,x\, q(x)$ populating the path-integral of the theory. The relevant
information is contained in the free energy density $f(\theta)$, which around
$\theta = 0$ can be usefully parametrized as a Taylor expansion as
follows~\cite{Vicari:2008jw}: 
\begin{equation}\label{eq:theta_dep}
f(\theta) = f(0) + \frac{1}{2}\chi\theta^2(1+b_2\theta^2+b_4\theta^4+\cdots)
\end{equation}
where the topological susceptibility $\chi$ and the coefficients
$b_{2n}$ can be related to the cumulants of the topological charge distribution
at $\theta=0$ by the relations 
\begin{equation}\label{eq:chi_b2}
\chi = \frac{\langle Q^2 \rangle_{c,\theta=0}}{\mathcal{V}}\ ,\quad
b_{2n}=
(-1)^n\, 
\frac{2\, \langle Q^{2n+2}\rangle_{c,\theta=0}}{(2n + 2)! \langle Q^2\rangle_{c,\theta=0}}\ ,
\end{equation}
where $\mathcal{V}$ is the four-dimensional volume. 

General large-$N$ arguments~\cite{Hooft-74, Witten-80, Witten-98} predict that,
in the low temperature confined phase of the theory, the susceptibility stays
finite in the large-$N$ limit, while the $b_{2n}$ are suppressed by increasing
powers of $1/N$, as follows;
\begin{equation}
\chi = \chi_\infty + O(N^{-2}), \quad  b_{2j}=O(N^{-2j}) \, .
\label{lnasyt0}
\end{equation}
Such predictions have been checked successfully both for
$\chi$~\cite{Lucini:2001ej,DelDebbio:2002xa,LTW-05,Ce:2016awn}, with
$\chi_\infty$ turning out to be compatible with the value predicted by the
Witten-Veneziano solution to the $U_A(1)$ problem~\cite{Witten-79,
Veneziano-79}, and for the fourth order coefficient
$b_2$~\cite{DelDebbio:2002xa,DElia:2003zne,Giusti:2007tu,Panagopoulos:2011rb,
Ce:2015qha,Bonati:2015sqt,Bonati:2016tvi}.

On the other hand, at asymptotically large $T$, i.e.~small compactification
radius, the theory becomes weakly coupled and one expects that instanton
calculus can be safely applied, leading to the validity of the dilute instanton
gas approximation (DIGA)~\cite{Gross:1980br, CDG-78}
\begin{eqnarray}
f(\theta) - f(0) &\simeq& \chi(T) \left( 1 - \cos\theta\right) \nonumber \\
\chi(T) &\simeq&  T^4 \exp[-8\pi^2/g^2(T)] \sim  T^{-\frac{11}{3} N + 4}, 
\end{eqnarray}
which predicts that the topological susceptibility vanishes exponentially fast
with $N$, while the $b_{2n}$ coefficients stay constant (for instance $b_2 =
-1/12$), contrary to the large-$N$ low-$T$ scaling.  The asymptotically large
temperature at which DIGA should set in is not known apriori; moreover, while
the prediction for $\chi (T)$ comes from a 1-loop computation, the $(1 - \cos
\theta)$ dependence expresses the fact that instantons and anti-instantons can
be treated as independent, non-interacting objects, which is the essential
feature of DIGA, and this could be true far before perturbative estimates
become reliable.

In fact, various theoretical arguments~\cite{KPT-98,BL-07,PZ-08} support the
idea that the change of regime should take place right after $T_c$, and faster
and faster as $N$ increases. This scenario is strongly supported by lattice
computations: the topological susceptibility drops at
$T_c$~\cite{susc_ft,GHS-02,DPV-04,LTW-05,Berkowitz:2015aua,Borsanyi:2015cka},
and it does so faster and faster as $N$ increases, pointing to a vanishing of
$\chi$ right after $T_c$ in the large-$N$ limit~\cite{DPV-04,LTW-05}. The
vanishing of $\chi$ might not be enough to prove that DIGA sets
in\footnote{There are various examples of quantum field theories with
non-trivial $\theta$-dependence where $\chi$ is predicted to vanish in some
limit, while the $b_{2n}$ coefficients do not reach their DIGA values, like
$CP^{N-1}$ models in two dimensions and in the large-$N$ limit
\cite{Rossi:2016uce, Bonati:2016tvi, Bonanno:2018xtd, Berni:2019bch} or QCD with dynamical
fermions in the chiral limit \cite{DiVecchia:1980yfw, diCortona:2015ldu,
Bonati:2015vqz}.}, so that a stronger and definite evidence comes from studies
of the coefficient $b_2$, proving that it reaches its DIGA value right after
$T_c$, and faster and faster as $N$
increases~\cite{Bonati:2013tt,Borsanyi:2015cka}.

As a consequence of the drastic change in the $\theta$-dependent part of the
free energy around $T_c$, the critical temperature itself is affected by the
introduction of a non-zero $\theta$, in particular $T_c$ turns out to be a
decreasing function of $\theta$~\cite{unsal-2,DElia:2012pvq,DElia:2013uaf}. 

The facts summarized above point to a strict relation between the
realization of center symmetry and the $\theta$-dependence of $SU(N)$
Yang-Mills theories, which one would like to investigate more closely. A
powerful tool, in this respect, is represented by trace deformed Yang-Mills
theories, which have been introduced in Ref.~\cite{Unsal:2008ch}, although
already explored by lattice simulations in Ref.~\cite{Myers:2007vc}.  The idea,
which is inspired by the perturbative form of the Polyakov loop effective
action at high temperature~\cite{Gross:1980br}, is to introduce one or more
(depending on the gauge group) center symmetric couplings to the Polyakov loop
and its powers, so as to inhibit the spontaneous breaking of center symmetry
even in the presence of an arbitrarily small compactification radius.  In this
way, one can approach the weak coupling regime, where semiclassical approaches
are available, while keeping center symmetry intact, so that the relation with
$\theta$-dependence can be investigated more systematically\footnote{Of course,
this offers the possibility to investigate to connection of center symmetry to
many other non-perturbative features of Yang-Mills theory, although in the
present study we are exclusively concerned with $\theta$-dependence.}.

Several works have already considered the use of trace deformed theories and
also possible alternatives, like the introduction of adjoint fermions or the
use of non-thermal boundary conditions~\cite{Kovtun:2007py, Unsal:2007vu,
Unsal:2007jx, Shifman:2008ja, Myers:2009df, Cossu:2009sq,  Meisinger:2009ne,
Unsal:2010qh, Thomas:2012ib, Poppitz:2012sw, Thomas:2012tu, Poppitz:2012nz,
Misumi:2014raa, Anber:2014lba, Bhoonah:2014gpa, Cherman:2016vpt,
Sulejmanpasic:2016llc, Anber:2017rch, Tanizaki:2019rbk, Itou:2018wkm, Bergner:2018unx}.  There are actually already well
definite semiclassical predictions regarding $\theta$-dependence in the
center-symmetric phase~\cite{Unsal:2008ch, Thomas:2011ee, unsal-1,
Aitken:2018mbb}, which come essentially from the fact that in the limit of
small compactification radius the deformed theory can be described in terms of
non-interacting objects with topological charge $1/N$ (a sort of Dilute
Fractional Instanton Gas Approximation, or DFIGA).  
This leads to predict $f(\theta) -
f(0) \propto 1 - \cos (\theta/N)$, hence for instance $b_2 = -1/ (12 N^2)$.
While these predictions are in agreement with general large-$N$ scaling for the
confined phase exposed above, they are not in quantitative agreement with the
lattice results for the confined phase, which yield instead $b_2 =
-0.23(3)/N^2$~\cite{Bonati:2016tvi}; in addition, also the topological
susceptibility itself is predicted to show significant deviations, for large
$N$ and small compactification radius~\cite{Thomas:2011ee}, from the behavior
showed in the standard confined phase.

It is therefore quite remarkable that, instead, lattice results obtained for
$SU(3)$, which have been reported for the first time in Ref.~\cite{gufo}, show
that one recovers exactly the same $\theta$-dependence as in the confined phase
(i.e.~the same value, within errors, for both $\chi$ and $b_2$) as soon as the
trace deformation is strong enough to inhibit the breaking of center symmetry.
The disagreement with semiclassical predictions is not a surprise, since the
values of the compactification radius $L$ explored in Ref.~\cite{gufo} go up to
$L^{-1} \equiv T \approx 500$~MeV, while the condition for the validity of the
semiclassical approximation is $T \gg N \Lambda$ where $\Lambda$ is the
non-perturbative scale of the theory, so that $T \sim 500$ MeV is a scale
where non-perturbative corrections can still be important. What is a surprise,
claiming for further investigations, is the fact that such non-perturbative
corrections are exactly the same as in the standard confined phase, leading to
the same $\theta$-dependence also from a quantitative point of view.

The purpose of the present study is to make progress along this line of
investigation, by extending the results of Ref.~\cite{gufo} to larger $SU(N)$
gauge groups, considering in particular the case $N = 4$.  There are various
reasons to expect that the study of $SU(4)$ may lead to new non-trivial
insights.  Apart from the fact that the space of trace deformations extends to
two independent couplings, we have that the possible breaking patterns of the
center symmetry group $\mathbb{Z}_4$ are more complex, including also a partial $\mathbb{Z}_4 \to
\mathbb{Z}_2$ breaking which corresponds to a phase differing from both the standard
confined and the deconfined phase of the undeformed theory. 

The way one can move across the different phases by tuning the two deformation
couplings can be predicted based on the 1-loop Polyakov loop effective
potential. However, as we will discuss, numerical simulations show the presence
of non-trivial corrections induced by fluctuations, which lead to complete
center symmetry restoration also when this is not expected. Morover, one has
the possibility to check whether the $\theta$-dependence of the standard
confined phase is achieved just for complete or also after partial restoration
of center symmetry.

The paper is organized as follows. In Section~\ref{setup} we review the
definition of $SU(N)$ pure gauge theories in the presence of trace
deformations, our lattice implementations and the numerical strategies adopted
to investigate $\theta$-dependence; in Section~\ref{results} we first compare
the predictions of 1-loop computations of the phase diagram with numerical
results, then discuss the $\theta$-dependence observed for the various
phases; finally, in Section~\ref{conclusions}, we draw our conclusions.

\section{Technical and numerical setup} 
\label{setup}

To investigate the relation between center symmetry and $\theta$-dependence we
will use, as already anticipated in Section~\ref{intro}, trace deformed Yang-Mills
theories. In order to inhibit the spontaneous breaking of center
symmetry when the theory is defined on a manifold with a compactified
dimension, new terms (the trace deformations) are added to the standard
Yang-Mills action, which are directly related to traces of powers of Polyakov
loops along the compactified direction. 

The action of the trace deformed $SU(N)$ Yang-Mills theory is thus \cite{Unsal:2008ch} 
\begin{equation}\label{tracedef_sun}
S^{\mathrm{def}} = S_{YM} + \sum_{\vec{n}} \sum_{j=1}^{\lfloor N/2\rfloor} h_j |\Tr P^{j}(\vec n)|^2 \ ,
\end{equation}
where $\vec{n}$ denotes a generic point on a surface orthogonal to the
compactified direction, the $h_j$s are new coupling constants, $P(\vec{n})$ is
the Polyakov loop associated to the compactified direction and $\lfloor \quad
\rfloor$ denotes the floor function. 
The number of possible trace deformations
is equal to the number of independent, center-symmetric functions of the 
Polyakov loop; in general, for $N > 3$, more than
one deformation could be needed, in order
to prevent the possibility of a partial
breaking of the center symmetry, 
with a nontrivial subgroup of $\mathbb{Z}_N$ left unbroken.

In order to clarify this point, let us specialize to the case $N=4$,
which is the one that will be thoroughly investigated in the following, and it
is the simplest case in which a partial breaking of center symmetry can 
take place.
For $N=4$ the action in Eq.~\eqref{tracedef_sun} reduces to 
\begin{equation}
S^{\mathrm{def}} = S_{YM} + h_1 \sum_{\vec n} |\Tr P(\vec n)|^2 
+ h_2 \sum_{\vec n} |\Tr P^2(\vec n)|^2 
\label{tracedef_su4}
\end{equation}
and complete restoration of $\mathbb{Z}_4$ requires the vanishing of the expectation
values of the two traces, $\Tr P$ and $\Tr P^2$. A priori none of the two new
terms in the action is sufficient to guarantee complete center symmetry
restoration: for instance, $M={\rm diag} (1,1,-1-1)$ has $\Tr M = 0$ but $\Tr
M^2 \neq 0$, while $M={\rm diag}(1,1,i,-i)$ has $\Tr M^2 = 0$ but $\Tr M \neq
0$. If $\langle \Tr P\rangle =0$ and $\langle \Tr P^2\rangle \neq 0$ (a
possibility which is forbidden if $N\le 3$) center symmetry is spontaneously
broken with the breaking pattern $\mathbb{Z}_4\to \mathbb{Z}_2$, which corresponds to the fact that single
quarks are confined but couples of quarks are not.

It thus seems that the term $|\Tr P(\vec{n})|^2$ in the action is needed to force
$\langle \Tr P\rangle=0$ and the term $|\Tr P^2(\vec{n})|^2$ to force $\langle
\Tr P^2\rangle=0$, but one should also take into account the following 
fact.
Trace deformations are spatially local quantities, i.e.\ they tend to
suppress $\Tr P(\vec{n})$ and $\Tr P^2(\vec{n})$ pointwise. However, the
restoration of a global symmetry can also be induced by disorder, since order
parameters are spatially averaged quantities, and this is what actually happens
in many well known cases, just like ordinary Yang-Mills theory (see, e.g., the
discussion on the adjoint Polyakov loop in Ref.~\cite{gufo}).
This will be particularly important in the following, when we will present an
analysis of the predicted phase diagram of the deformed $SU(4)$ gauge theory
based on the 1-loop effective potential of the Polyakov loop: this kind of
analysis assumes a spatially uniform Polyakov loop, hence neglects the
possibility of long-distance disorder. This is a possible explanation of the
fact that numerical results will show sometimes deviations from the 1-loop
effective potential prediction, so that, for instance, center symmetry can be
restored completely in some cases by adding just one trace.

The discretization of the action in Eq.~\eqref{tracedef_su4} does not present
particular difficulties: for the Yang-Mills action $S_{YM}$ we adopt the standard Wilson
action \cite{Wilson:1974sk} (in the following $\beta$ will denote the bare
coupling $\beta=6/g^2$) and trace deformations can be rewritten
straightforwardly in terms of the lattice variables. The update of the links directed along
spatial directions can be performed by using heatbath and overrelaxation algorithms
\cite{Creutz:1980zw, Kennedy:1985nu, Creutz:1987xi} implemented \emph{\`a la}
Cabibbo-Marinari \cite{Cabibbo:1982zn}, while for the temporal links (which do not 
enter linearly in the action) we have to resort to a
Metropolis update \cite{Metropolis:1953am}. 
\\

The procedure we used to assign an integer topological charge value $Q$ to a
given configuration is the following \cite{DelDebbio:2002xa}: first of all
we reduced the ultraviolet noise present in the configuration by using
cooling~\cite{Berg:1981nw, Iwasaki:1983bv, Itoh:1984pr, Teper:1985rb,
Ilgenfritz:1985dz}  (the numerical equivalence of different smoothing
algorithms was shown in several studies, see Refs.~\cite{Bonati:2014tqa,
Cichy:2014qta, Namekawa:2015wua, Alexandrou:2015yba, Berg:2016wfw,
Alexandrou:2017hqw}), then we computed on the smoothed configurations the
quantity $Q_{ni}=\sum_x q_L(x)$, where $q_L(x)$ is the discretization
of the topological charge density introduced in Refs.~\cite{DiVecchia:1981qi,
DiVecchia:1981hh}
\begin{equation}\label{eq:qlattice}
q_L(x) = -\frac{1}{2^9 \pi^2} 
\sum_{\mu\nu\alpha\beta = \pm 1}^{\pm 4} 
{\tilde{\epsilon}}_{\mu\nu\alpha\beta} \hbox{Tr} \left( 
\Pi_{\mu\nu}(x) \Pi_{\alpha\beta}(x) \right) \; .
\end{equation}
In this expression $\Pi_{\mu\nu}$ is the plaquette operator and the modified
Levi-Civita tensor $\tilde{\epsilon}_{\mu\nu\alpha\beta}$ coincides with the
standard one for positive indices, while its value for negative indices is
completely determined by ${\tilde{\epsilon}}_{\mu\nu\alpha\beta} =
-{\tilde{\epsilon}}_{(-\mu)\nu\alpha\beta}$ and complete antisymmetry.  The
integer value of the topological charge $Q$ is finally related to $Q_{ni}$ by
\begin{equation}
Q=\mathrm{round}(\alpha Q_{ni})\ ,
\end{equation}
where ``$\mathrm{round}$'' stands for the rounding to the closest integer and the
constant $\alpha$ was fixed in such a way as to make $\langle (Q-\alpha
Q_{ni})^2\rangle$ as small as possible 
{(see Refs.~\cite{DelDebbio:2002xa,Bonati:2015sqt} for more details).}

From the Monte-Carlo history of $Q$ it is straightforward to estimate the
topological susceptibility by using Eq.~\eqref{eq:chi_b2}. This is a priori
possible also for the coefficient $b_2$, however this is known not to be the
most efficient way of extracting it: a $b_2$ estimator with a more favorable
signal-to-noise ratio (especially for large volumes) can be obtained by
performing simulations at non-vanishing (imaginary, to avoid the sign problem)
values of the topological $\theta$ angle~\cite{Panagopoulos:2011rb,
Bonati:2015sqt, Bonati:2016tvi}. 

In practice, if a $\theta$-term of the form $-\theta_L q_L(x)$ is added to the
lattice action, $b_2$, $\chi$  and the finite lattice renormalization constant
of $q_L(x)$ \cite{Campostrini:1988cy} can be extracted from the cumulants of
the topological charge distribution at $\theta_L\neq 0$. This approach,
although apparently more computationally demanding than the standard one at
$\theta_L=0$, turns out in fact to be much more efficient to obtain reliable
estimates of $b_2$.  For more details we
refer to Ref.\cite{Bonati:2015sqt}, where the same method used in the present study
was adopted and explained at length. 

We finally note that, despite the advantages of the imaginary-$\theta$ method,
a determination of $b_2$ is still significantly more challenging than a
determination of the topological susceptibility. For this reason in
Section~\ref{sec:theta} we will use the topological susceptibility when performing
a broad scan of the $\theta$-dependence across the phase diagram, while $b_2$
will be measured only for some specific points.

\section{Results} 
\label{results}

The description of our numerical results is divided in two steps. First,
we will discuss the phase structure of the deformed $SU(4)$ gauge theory 
in the $h_1$-$h_2$ plane and for values of the compactification radius 
(temperature) for which center symmetry is broken at $h_1 = h_2 = 0$:
we will make use of predictions coming from the 1-loop effective potential,
and compare them with results from numerical simulations. In the second part,
the $\theta$-dependence which is found in the different phases will be
presented and discussed.

\begin{figure}[t]
\includegraphics[width=0.699\columnwidth, clip]{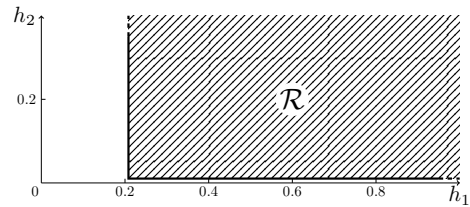}
\caption{Graphical representation, in the plane $(h_1,h_2)$, of the region $\mathcal{R}$ 
corresponding to points for which $\lambda_k=e^{i\alpha_k}$ ($k=0,\ldots,3$),
with $\alpha_k=\frac{\pi}{4}+k\frac{\pi}{2}$, is a local minimum of the 1-loop
effective potential.}
\label{stabilityregion}
\end{figure}

\subsection{Phase diagram inthe deformation space: 1-loop predictions confront
numerical results}
\label{sec:phase}

\begin{figure}[t]
    \includegraphics[width=0.8\columnwidth, clip]{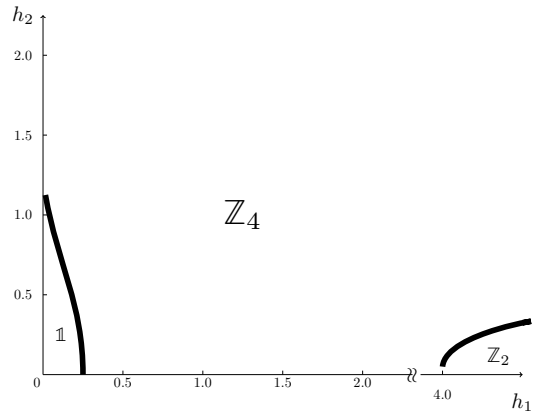}
\caption{Phase diagram obtained from simulations performed at bare coupling
$\beta = 11.15$ on a $6 \times 32^3$ lattice, corresponding to 
an inverse compactification radius $L^{-1} = T \simeq 393$ MeV.}
\label{beta1115pd}
\end{figure}

In the perturbative regime, the effective potential of a translation invariant
$SU(4)$ configuration (with $P(\vec{n})\equiv P$) assumes the form
\cite{Unsal:2008ch}
\begin{equation} \label{1looppotential}
V[P] = \mathcal{E}(P) + h_1 |\Tr P|^2 +  h_2 |\Tr P^2|^2\ ,
\end{equation}
where $\mathcal{E}(P)$ is the 1-loop effective potential of the standard Yang-Mills theory
computed in Ref. \cite{Gross:1980br}:
\begin{equation}\label{GPYpot}
\mathcal{E}(P) = \sum_{k = 1}^{\infty} \frac{1}{k^4} |\Tr P^k|^2\ .
\end{equation}
Since Eq.~\eqref{1looppotential} is an $SU(4)$ invariant function, the
effective potential can be conveniently rewritten as a function of the three
independent eigenvalues of $P$. 

Despite the apparent simplicity of Eq.~\eqref{1looppotential}, it is far from
trivial to obtain a closed analytical expression for the position of its absolute
minimum. It is nevertheless possible to gain some analytical insight into the
breaking of center symmetry and the structure of the phase diagram of the
$SU(4)$ deformed Yang-Mills theory. Every matrix $M\in SU(4)$ satisfying $\Tr
M=\Tr M^2=0$ is equivalent to the diagonal matrix with eigenvalues
$\lambda_k=e^{i\alpha_k}$ ($k=0,\ldots,3$), with
$\alpha_k=\frac{\pi}{4}+k\frac{\pi}{2}$. If we denote by $\mathcal{R}$ the
region of the $(h_1, h_2)$ plane corresponding to points for which
$\{\lambda_k\}$ is a local minimum of Eq.~\eqref{1looppotential}, the parameter
region in which center symmetry is not broken is necessarily a subset of
$\mathcal{R}$ and $\mathbb{Z}_4$ is surely broken for all the values $(h_1,h_2)$ outside
$\mathcal{R}$. The region $\mathcal{R}$ can be analytically determined and it
can be seen that 
\begin{equation}
\mathcal{R}=\{h_1>\frac{5}{24}\}\cap \{h_2>\frac{1}{96}\}\ ,
\end{equation}
as shown in Fig.~\ref{stabilityregion}.  In particular, as anticipated, we see
that a single deformation is not sufficient to ensure the absence of center
symmetry breaking in the 1-loop effective action: the axes $h_1=0$ and $h_2=0$
lay outside $\mathcal{R}$ and $\mathbb{Z}_4$ has to be broken there.

To test the effectiveness of the 1-loop potential in predicting the phase
diagram, we also numerically investigate the phase diagram of the lattice
deformed Yang-Mills theory, using a $6\times 32^3$ lattice and two values of
the lattice coupling larger than the critical value $\beta_c\simeq 10.79$ (see
Ref.~\cite{Lucini:2005vg}). More in detail, we considered $\beta = 11.15$ 
(corresponding to
an inverse compactification radius $T \approx 393$ MeV)
and $\beta = 11.40$ ($T
\approx 482$ MeV), then performed a scan of the plane $(h_1,h_2)$ in the range
$[0,2] \times [0,2]$ with a step $\Delta=0.1$, for a total of $441$ simulation
points for each $\beta$ value.
The scale has been fixed using the determination of 
Ref.~\cite{Lucini:2005vg} (see in particular Eq.~(35) therein) 
and fixing the string tension to be
$\sigma = (440 \ \mathrm{MeV})^2$.

\begin{figure}[t]
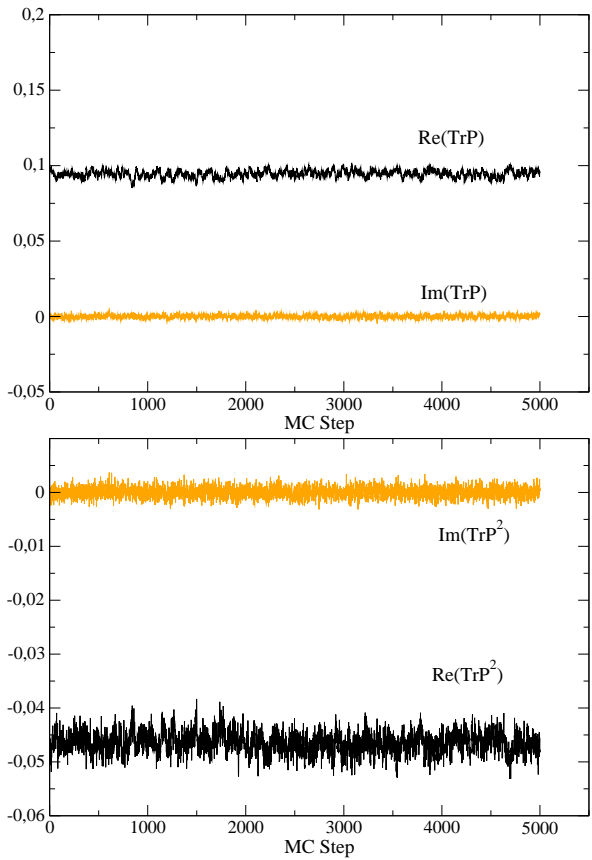

\includegraphics[width=0.9\columnwidth, clip]{repimp.eps}
\includegraphics[width=0.9\columnwidth, clip]{repimp2.eps}
\caption{An example of complete breaking of center symmetry ($\mathbb{Z}_4\to\mathrm{Id}$) at $\beta = 11.15$. 
We report the Monte-Carlo histories
of $\mathrm{Re}(\Tr P)$, $\mathrm{Im}(\Tr P)$, $\mathrm{Re}(\Tr P^2)$,
$\mathrm{Im}(\Tr P^2)$ for $h_1 = 0.0$, $h_2=0.1$. Both $\mathrm{Re}(\Tr P)$ and $\mathrm{Re}(\Tr P^2)$ are non-zero.}
\label{z4broken}
\end{figure}

\begin{figure}[t]
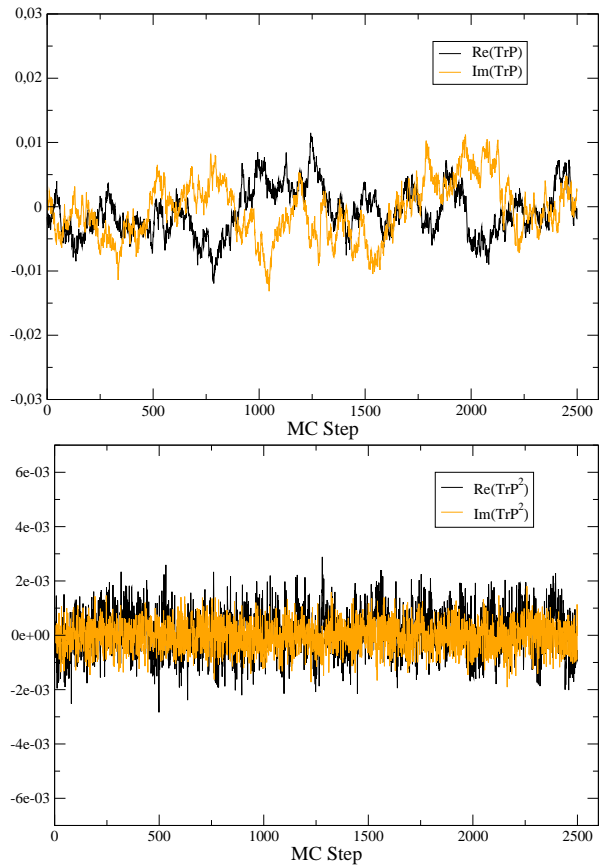

\includegraphics[width=0.91\columnwidth, clip]{reim_017.eps}
\includegraphics[width=0.91\columnwidth, clip]{reim2_017.eps}
\caption{An example of complete restoration 
of center symmetry for $\beta = 11.15$, $h_1 = 0.0$ and $h_2=1.7$.  
The Monte-Carlo histories of all quantities, 
$\mathrm{Re}(\Tr P)$, $\mathrm{Im}(\Tr P)$, $\mathrm{Re}(\Tr P^2)$, and
$\mathrm{Im}(\Tr P^2)$, fluctuate around their zero average values.
It is interesting to notice that the fluctuations of $\Tr P$
are significantly larger than those of $\Tr P^2$: indeed 
$\langle P \rangle$ should not 
be zero according to the 1-loop effective potential, and
vanishes because of long range disorder.}
\label{z4safe}
\end{figure}

\begin{figure}[h!!]
\includegraphics[width=0.9\columnwidth, clip]{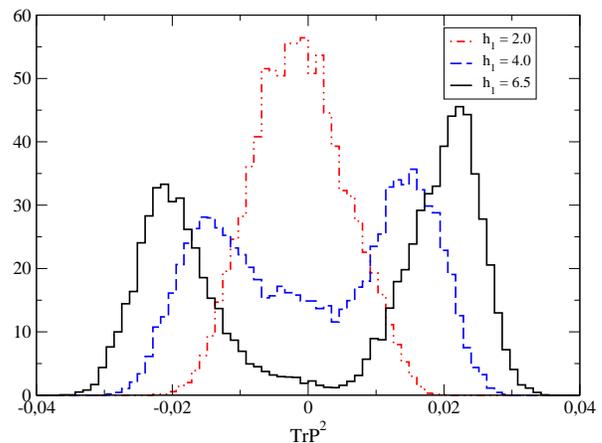}
\caption{The histogram of $\mathrm{Re}(\mathrm{Tr} P^2)$ computed using
a $6 \times 32^3$ lattice at bare coupling $\beta = 11.15$ for three different values
of $h_1$ along the $h_2 = 0$ axis. }
\label{fig:histogram_h1_vs_0}
\end{figure}

\begin{figure}[t]
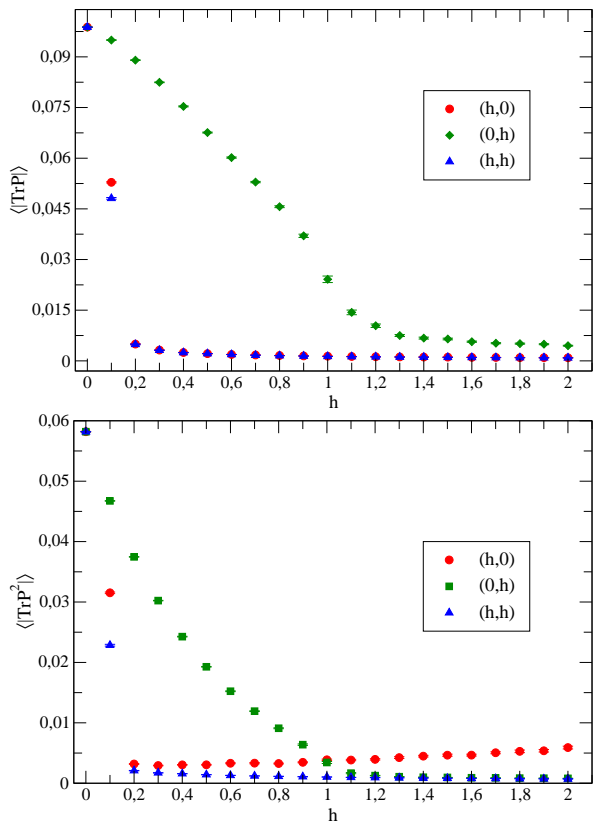

\includegraphics[width=0.9\columnwidth, clip]{modP_vs_h_1115.eps}
\includegraphics[width=0.9\columnwidth, clip]{modP2_vs_h_1115.eps}
\caption{$\langle |\Tr P|\rangle$  and $\langle |\Tr P^2| \rangle$ computed using
a $6\times 32^3 $ lattice at bare coupling $\beta = 11.15$ for different values
of the deformation parameters. Different datasets correspond to deformations of
the form ($h_1 \neq 0$, $h_2=0$), ($h_1 = 0$, $h_2 \neq 0 $) and ($h_1 = h_2$).}
\label{fig:b1115_modP}
\end{figure}

\begin{figure}[h!!]
\includegraphics[width=0.9\columnwidth, clip]{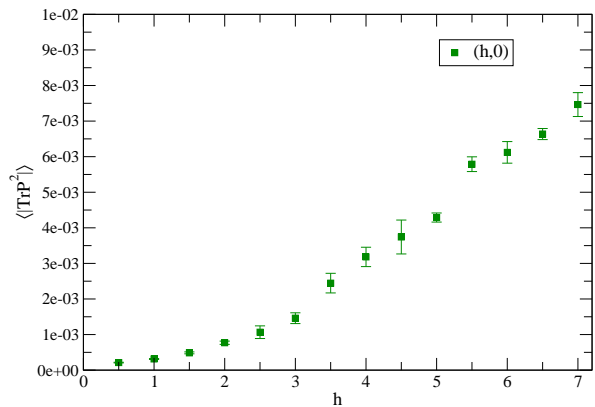}
\caption{$\langle |\Tr P^2| \rangle$ computed using
a $6 \times 32^3$ lattice at bare coupling $\beta = 11.15$ for different values
of $h_1$ along the $h_2 = 0$ axis.}
\label{fig:b1115_modP_extended}
\end{figure}

The phase diagram obtained from numerical simulations performed at $\beta =
11.15$ is shown in Fig.~\ref{beta1115pd}: 
in a small region around the origin
$\mathbb{Z}_4$ is completely broken, while outside there is no breaking at all, apart from a region at 
large values
of $h_1$, where $\mathbb{Z}_4$ breaks partially to $\mathbb{Z}_2$.
The different phases have been identified both by looking
at histograms of the time-histories of $\Tr P$, $\Tr P^2$ 
(see Figs.~\ref{z4broken}, ~\ref{z4safe} and \ref{fig:histogram_h1_vs_0}) 
and by studying\footnote{Note that $\langle \Tr P\rangle$
and $\langle \Tr P^2\rangle$ identically vanish on finite lattices, apart from
possible numerical issues related to ergodicity breaking for large volumes.}
$\langle |\Tr P|\rangle$, $\langle |\Tr P^2|\rangle$ 
(see Figs.~\ref{fig:b1115_modP} and \ref{fig:b1115_modP_extended}), where
\begin{equation}
P\equiv \frac{1}{\mathcal{V}}\sum_{\vec{n}}P(\vec{n}) \, .
\end{equation}
The picture that emerges is in
striking contrast with the expectations based on the 1-loop effective
potential: even a single deformation is capable of completely stabilizing
center symmetry ($0.2 < h_1 \lesssim  4$  for $h_2=0$, or 
$h_2 > 1.1$ for $h_1 = 0$). This can be noticed by looking at Fig.~~\ref{fig:histogram_h1_vs_0}
and Fig.\ref{fig:b1115_modP_extended}. 

\begin{figure}[t]
\includegraphics[width=0.7\columnwidth, clip]{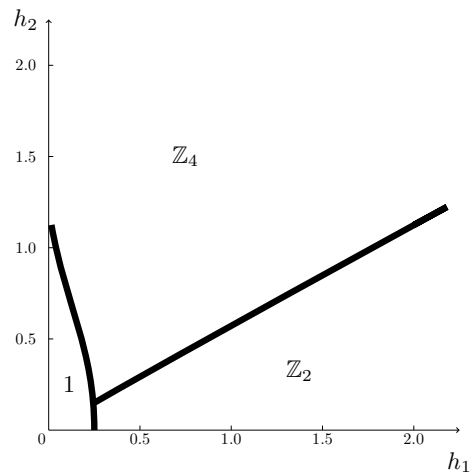}
\caption{Phase diagram obtained from simulations performed at bare coupling
$\beta = 11.40$ on a $6 \times 32^3$ lattice, corresponding to 
an inverse compactification radius $L^{-1} = T \simeq 482$ MeV.}
\label{beta1140pd}
\end{figure}

Moving to the larger value of $\beta$ that 
we have explored (corresponding 
to a smaller compactification radius), one may expect that
predictions based on the 1-loop effective
potential get more reliable.
The phase diagram
obtained for $\beta=11.40$ is shown in Fig.~\ref{beta1140pd}. 
We can see that indeed the new partially broken phase becomes more 
manifest, so that center symmetry is now broken along the whole
$h_1$ axis, as predicted in terms of the 1-loop potential;
however, along the $h_1=0$ axis the discrepancy 
persists, with center symmetry being protected just by the 
$|\Tr P^2(\vec{n})|^2$ deformation.

Notice that in sketching Fig.~\ref{beta1140pd} we have not made any statement about the 
order of the various transition lines. This is an issue that should be considered in future studies and by now we can just 
make some general statements: direct transitions from the completely broken phase to the completely restored phase are expected 
to be first order, as for the standard deconfining phase transition of $SU(4)$, while transition from the partially restored phase should 
be in the universality class of the 3D Ising model if they are second order, however they can still be first order, this depends on the dynamics
of the system and should be checked by more extensive numerical simulations.

To further investigate the origin of the inconsistencies between the prediction of the
1-loop effective potential and the phase diagram observed in numerical
simulations, we studied the quantities 
\begin{gather}\label{localquantities}
\langle |\Tr P_{loc}|^2\rangle\equiv \frac{1}{\mathcal{V}} \sum_{\vec n} \langle |\Tr P(\vec n)|^2  \rangle   \\
\langle |\Tr P_{loc}^2|^2\rangle\equiv \frac{1}{\mathcal{V}} \sum_{\vec n} \langle |\Tr P^2(\vec n)|^2\rangle  \ .
\end{gather}
Since the squared modulus in this case is taken over local, rather than
spatially averaged, quantities, such observables should be less sensitive
to long range disorder and follow more closely the prediction of the 
1-loop effective potential.

\begin{figure}[t]
	\includegraphics[width=0.9\columnwidth, clip]{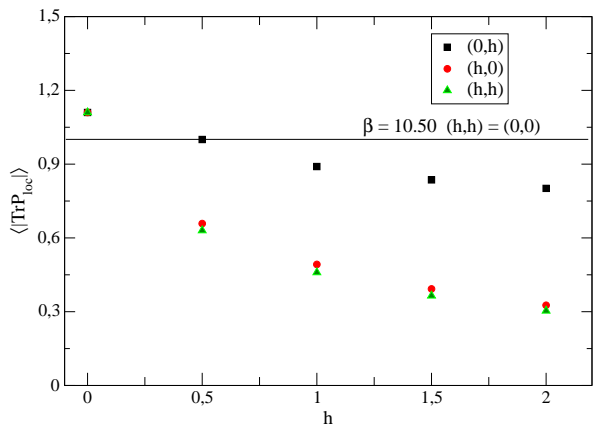}
	\caption{Mean value of the local quantity $\langle |\Tr P_{loc}|^2\rangle$.
	    The black line indicates the value of the undeformed theory for $\beta = 10.50$.
      The lattice used is $6\times 32^3$ and the bare coupling $\beta = 11.15$.}
	\label{fig:polylocal}
\end{figure}

\begin{figure}[t]
	\includegraphics[width=0.9\columnwidth, clip]{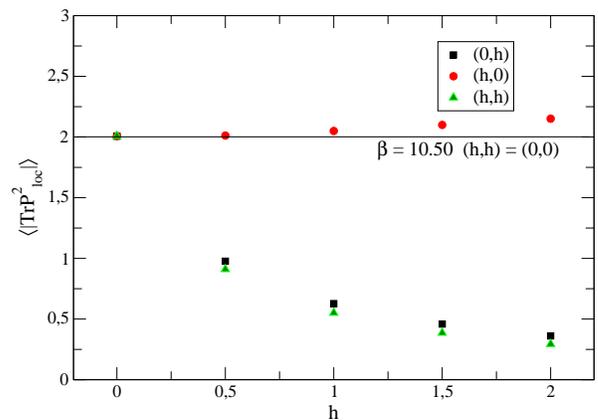}
	\caption{Mean value of the local quantity $\langle |\Tr P_{loc}^2|^2\rangle$.
	 The black line indicates the value of the undeformed theory for $\beta = 10.50$.
   The lattice used is $6\times 32^3$ and the bare coupling $\beta = 11.15$.}
	\label{fig:poly2local}
\end{figure}

Our results have been obtained by performing simulations using three
different setups for the deformation parameters $h_1$ and $h_2$ in
Eq.~\eqref{tracedef_su4}: the first two setups are the ones in which only a single
deformation is present, i.e. $h_1 \neq 0 $ and $h_2=0$ or $h_1 = 0 $ and $h_2
\neq 0$. The third setup is the one in which both  deformations are active
and, for the sake of the simplicity, we restricted to the ``diagonal'' configuration
$h_1=h_2$. 
We show in particular results obtained 
for $\beta = 11.15$ on the $6 \times 32^3$ lattice 
(which is one of the two setups already discussed above),
which are reported in Figs.~\ref{fig:polylocal} and~\ref{fig:poly2local}
and there compared to reference values obtained on the same lattice
and without any deformation at $\beta = 10.50$, 
which is deep into the confined phase. 
The corresponding quantities, for which the squared modulus is taken
after the spatial average, have been already shown in 
Fig.~\ref{fig:b1115_modP}.

The general lesson we can learn by comparing the different 
behaviors is the following. On one hand, it is clear that 
the local quantities,  $\langle |\Tr P_{loc}|^2\rangle$ and 
$\langle |\Tr P^2_{loc}|^2\rangle$, are significantly more suppressed,
with respect to their values in the standard confined phase, 
when a direct coupling to the relevant deformation is present
(i.e., respectively, $h_1 \neq 0$ or $h_2 \neq 0$); this fact
was already noticed and discussed in Ref.~\cite{gufo}, pointing 
out to a different kind (from a dynamical point of view) of center symmetry restoration 
in the trace deformed theory,
with respect to the standard confined
phase.

On the other hand, when no direct coupling to the relevant 
deformation is present (i.e., along the $(0,h)$ axis
for $\langle |\Tr P_{loc}|\rangle$ and along the $(h,0)$ axis
for $\langle |\Tr P_{loc}|^2\rangle$), the local quantities are
not significantly suppressed or remain almost constant, in agreement with 
the predictions of the 1-loop effective potential, meaning that
in this case the complete restoration of center symmetry 
takes place because of long range disorder. 
This is also appreciable from Fig.~\ref{z4safe},
where the Monte-Carlo histories of the spatially averaged
quantities are shown for the same $\beta$ value
and for a point along the $(0,h)$ axis where $\mathbb{Z}_4$ is 
completely restored: $\Tr P$, which is not coupled to any 
deformation, averages to zero, but with much larger fluctuations with respect
to $\Tr P^2$; we interpret this as a manifestation of the fact
that $\Tr P$ is locally non-zero, but fails to reach an ordered
phase at large scales.ì

\subsection{$\theta$-dependence of the various phases}
\label{sec:theta}

\begin{figure}[t]
\includegraphics[width=0.9\columnwidth, clip]{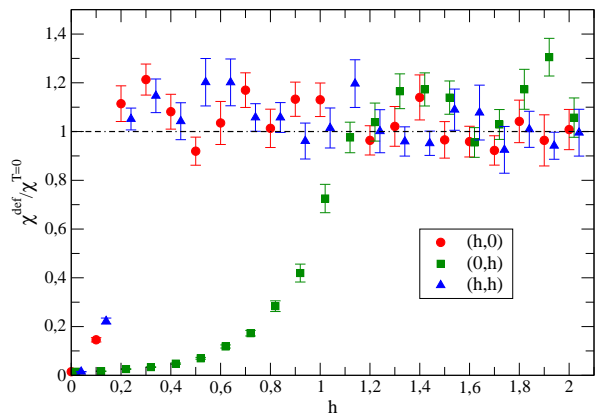}
\caption{Ratio between the topological susceptibility $\chi$ computed in the
deformed theory and the one at $T=0$ computed in 
Ref.~\cite{Bonati:2016tvi} for different values of the deformation parameters
$h_1$ and $h_2$. Results are obtained on the $6\times 32^3$ lattice 
at bare coupling 
$\beta = 11.15$.}
\label{fig:b1115_top_susc}
\end{figure}

We are now going to discuss the $\theta$-dependence of the
different phases identified previously for the deformed $SU(4)$
theory. It is interesting, in particular, 
to ask whether the different ways in which $\mathbb{Z}_4$ can
be restored manifest themselves also in a different 
$\theta$-dependence or not.
Let us start from the case of the $6\times 32^3$ lattice at bare coupling
$\beta = 11.15$ ($T\approx 393$ MeV), whose phase diagram was shown in
Fig.~\ref{beta1115pd}. In Fig.~\ref{fig:b1115_top_susc} we report the behaviour
of the topological susceptibility $\chi$ as a function of the deformation
parameters $h_1$ and $h_2$, for the three deformation setups introduced above.
In order to have a direct comparison with the $T=0$ result, 
we plot the ratio
between the topological susceptibility $\chi$ in the deformed theory and the
$T=0$ continuum  value computed in ordinary Yang-Mills 
theory in Ref.~\cite{Bonati:2016tvi}. We are using here the fact, 
explicitly verified in Ref.~\cite{gufo}, that the lattice spacing can be considered to be independent of
the deformation for all practical purposes. This will not be necessary in
the following when discussing results for $b_2$, since $b_2$ is dimensionless. 

For $h_1=h_2=0$ the system at $\beta=11.15$ is in the deconfined phase, so we
expect the value of the topological susceptibility $\chi$ to be tiny for small
values of the deformation parameters. From data in
Fig.~\ref{fig:b1115_top_susc} we see that this is indeed the case for all the
deformation setups studied. Moreover, the topological susceptibility always
reaches a plateau for large deformations, at a value which is consistent with
that of $\chi$ measured at $T=0$ in ordinary Yang-Mills theory. This
asymptotic value is however approached differently in the different deformation
setups: when using $h_2=0$ or $h_1=h_2$ the plateau starts from $h\approx 0.2$,
while in the setup with $h_1=0$ it starts from $h\approx 1.2$. The reason for
this behaviour is clear from the phase diagram shown in Fig.~\ref{beta1115pd}:
these values of the deformation parameters are the one that are needed to reach
the $\mathbb{Z}_4$-symmetric phase when moving along the axes or along the diagonal of
the phase diagram.

\begin{figure}[t]
\includegraphics[width=0.9\columnwidth, clip]{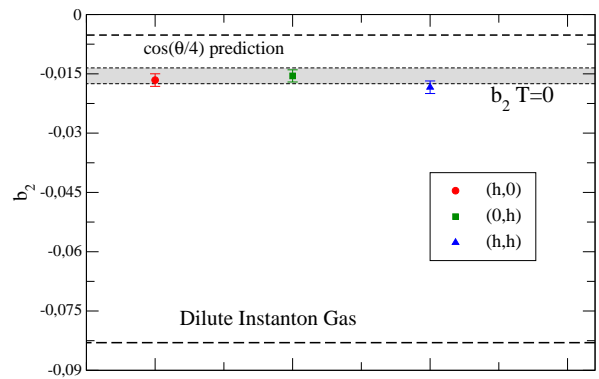}
\caption{The coefficient $b_2$ measured in the deformed theory on the $6\times
32^3$ lattice at $\beta = 11.15$, for the different deformation setups and $h=1.5$.  The
band denotes the $T=0$ continuum result computed in 
Ref.~\cite{Bonati:2016tvi}, while the dashed lines indicate the DIGA ($-1/12$)
and the DFIGA 
prediction ($-1/192$).}
\label{fig:b1115_b2}
\end{figure}

Using the same lattice setting we computed also the coefficient $b_2$ related
to the fourth power of $\theta$ in the expansion of the free energy, see
Eq.~\eqref{eq:theta_dep}.  As explained in Sec.~\ref{setup}, the estimation of
$b_2$ is computationally much more demanding than that of $\chi$; for this
reason we decided to compute $b_2$ just for three values of the deformations
deep in the plateau region, one for each of the three deformation setups
previously adopted (with $h=1.5$ in all the cases). We computed $b_2$ by means
of the imaginary $\theta$ method discussed in Sec.~\ref{setup}, using
7 values of $\theta_L$ in the range $[0,12]$.  The
outcome of this analysis is reported in Fig.~\ref{fig:b1115_b2}: also for $b_2$
there is a nice agreement between the values computed in the deformed theory in
the $\mathbb{Z}_4$ restored phase and the one obtained in the $T=0$ Yang-Mills case, for
all the deformation setups.

It is interesting to compare the results obtained for $b_2$ with the values
predicted by using two well known approximation schemes. The first one is the
DIGA, which is expected to be reliable in ordinary
Yang-Mills theory for a small value of the compactification radius. In this
approximation the system is supposed to be well approximated by a gas of weakly
interacting degrees of freedom, carrying an unit of topological charge ($\pm
1$), and the coefficient $b_2$ is predicted to be $-1/12$. The second
approximation scheme is the DFIGA,
which is expected to be valid in the center symmetric phase of the deformed
theory for small values of the compactification length.  In this case the
degrees of freedom are still expected to be weakly interacting, but now they
carry a fractional topological charge, quantized in units of $1/N$. In this
scenario the predicted value is $b_2 = -1/(12N^2)$, i.e.~$b_2 = -1/192$ for
$SU(4)$. Both these values are shown in  Fig.~\ref{fig:b1115_b2} and they are
clearly not compatible with numerical data, indicating that the
compactification length used is still too large for the interactions between
the fractional degrees of freedom to be negligible.

\begin{figure}[t]
\includegraphics[width=0.9\columnwidth, clip]{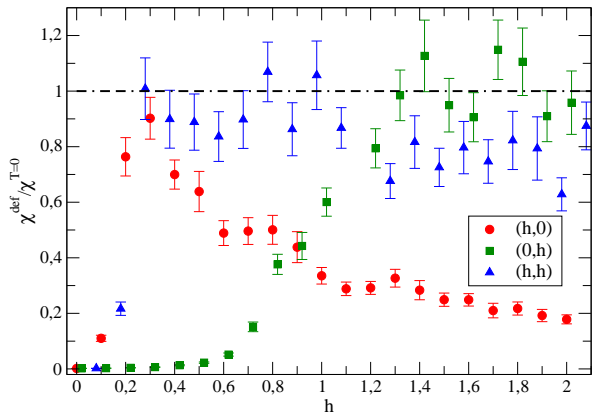}
\caption{Ratio between the topological susceptibility $\chi$ computed in the
deformed theory and at the one at $T=0$ continuum
from \cite{Bonati:2016tvi} for different values of the deformation parameters
$h_1$ and $h_2$. Results are obtained on the lattice $6\times 32^3$ at bare coupling 
$\beta = 11.40$.}
\label{fig:b1140_top_susc}
\end{figure}

Let us now repeat the same analysis for the second value of the bare coupling
constant $\beta$ studied in Section~\ref{sec:phase}, i.e. $\beta=11.40$ (corresponding to
$T\approx 482$ MeV). The values of the deformations used are $0\leq h_1 \leq2$ 
and $0\leq h_2 \leq 2$. Three different phases are present, see
Fig.~\ref{beta1140pd}, and one could expect that also the $\theta$-dependence
shows some signal of the presence of the phase with $\mathbb{Z}_4$ broken to $\mathbb{Z}_2$.

\begin{figure}[t]
\includegraphics[width=0.9\columnwidth, clip]{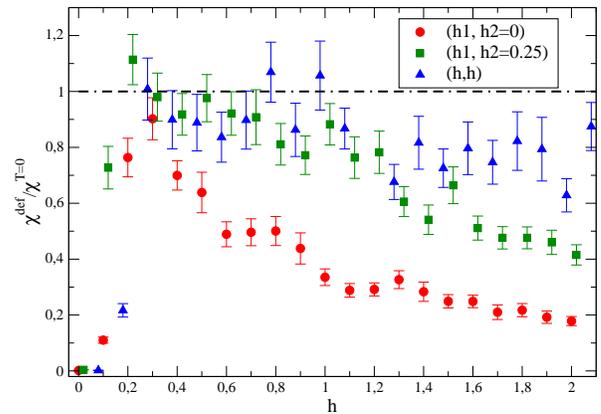}
\caption{Ratio between the topological susceptibility $\chi$ computed in the
deformed theory and at the one at $T=0$ continuum
from Ref.~\cite{Bonati:2016tvi} for different values of the deformation parameters
$h_1$ and $h_2$. In particular here we report the case
in which $h_1$ varies and $h_2$ is kept fixed 
at $h_2=0.25$. Results are obtained on the $6\times 32^3$ lattice 
at bare coupling 
$\beta = 11.40$.}
\label{fig:b1140_top_susc_partial}
\end{figure}

From Fig.~\ref{fig:b1140_top_susc}, where the results for the topological
susceptibility are reported, we see that this is indeed the case: errors are
larger than for $\beta=11.15$ but it is quite clear that the values of $\chi$
approach $\chi_{T=0}$ only for two of the three deformation setups adopted,
namely the one in which $h_1=0$ and the one in which $h_1=h_2$. By looking at
the phase diagram in Fig.~\ref{beta1140pd} we see that these are the only two
setups in which the deformations induce a complete restoration of the center
symmetry, and that the values of the deformation at which the plateaux are
reached are consistent with the boundaries of the region with broken center
symmetry. In the remaining deformation setup, in which $h_2=0$, center symmetry
is not completely restored by increasing the value of $h_1$, and the system
enters the phase in which center symmetry is broken to its $\mathbb{Z}_2$ subgroup. While
it is not clear why in this phase the susceptibility seems to approach zero as
we increase $h_1$, it is tempting to interpret the peak at $h\approx 0.3$ (at
which point $\chi\approx \chi_T$) as a proximity effect due to the closeness of
the completely restored phase in the phase diagram (see Fig.~\ref{beta1140pd}).
In order to investigate this hypothesis we computed the value of the topological
susceptibility also using a different setup, i.e. varing $h_1$ and keeping $h_2=0.25$,
because from the phase diagram of Fig.~\ref{beta1140pd} we see that in this setup the system passess
across all the symmetry breaking patterns. Results are shown in Fig.~\ref{fig:b1140_top_susc_partial}.
We can clearly see that the case $(h_1, h_2 = 0.25)$ is in between the ``diagonal" 
case and the one with only the $h_1$ deformation: the values of the deformation
parameter at which the topological susceptibility is compatible with the one at $T=0$
correspond to the region in which center symmetry is completely restored, this can be 
appreciated comparing with the phase diagram shown in Fig.~\ref{beta1140pd}.

\begin{figure}[t]
\includegraphics[width=0.9\columnwidth, clip]{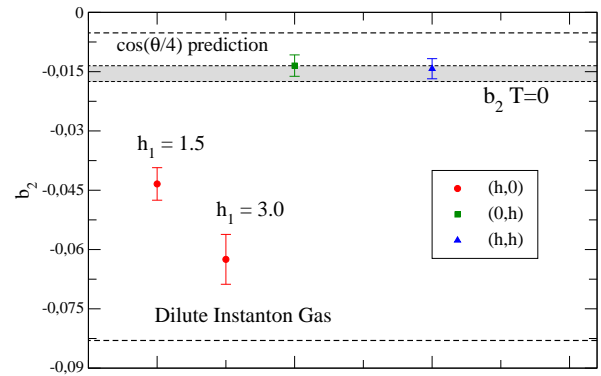}
\caption{The coefficient $b_2$ measured in the deformed theory on the $6\times
32^3$ lattice at $\beta = 11.40$, for the different deformation setups and
$h=1.5$ (apart from the point indicated by by $h_1=3.0$). The band denotes the
$T=0$ continuum result of Ref.~\cite{Bonati:2016tvi},
while the dashed lines indicate the DIGA ($-1/12$) and the DFIGA prediction ($-1/192$).}
\label{fig:b1140_b2}
\end{figure}

The presence of the partially broken phase is evident also from the values of
$b_2$ computed at $\beta=11.40$, which are shown in Fig.~\ref{fig:b1140_b2}.
The values of $b_2$ in the phase with completely restored center symmetry are
again compatible with the results obtained at $T=0$ in \cite{Bonati:2016tvi},
while the values corresponding to the deformation parameters $h_1=1.5, h_2=0$
and $h_1=3.0, h_2=0$ are incompatible with $b_2(T=0)$, and lay in the
middle between the DIGA prediction $-1/12$ and the DFIGA one $-1/192$.

Altogether lattice data indicate that the $\theta$-dependence of the deformed
theory concides with the one of ordinary Yang-Mills theory at $T=0$ only when
center symmetry is completely recovered, 
and this happens independently of the specific way the restoration
takes place, i.e.~either by local suppression of $\Tr P$ 
and $\Tr P^2$, or by long range disorder.
Instead, in the phase in which center symmetry
is only partially restored both the topological susceptibility and $b_2$ do not
reach a clear plateau as a function of the deformation parameter, and they
assume values somewhere in between the deconfined and the confined case.

\section{Conclusions} 
\label{conclusions}
In this paper we have investigated 
the relation between center symmetry
and $\theta$-dependence in
Yang-Mills theories, 
exploiting trace deformations in order to control the realization of 
center symmetry breaking in a theory with a small compactified direction.
Extending previous results presented in Ref.~\cite{gufo} for the 
$SU(3)$ pure gauge theory, we have considered $SU(4)$, which is particularly
interesting since, apart from allowing a larger space of independent
trace deformations, is also the first $SU(N)$ gauge group for which 
the center group admits various patterns of symmetry breaking.

As a first step, we have investigated the 
phase diagram of the theory in the deformation space and
for various values
of the inverse compactified radius, reaching values up
to $L^{-1} \sim 500$~MeV. We have considered predictions from the 
1-loop effective potential of the Polyakov loop and compared 
them to results of numerical lattice simulations, in which 
the fate of center symmetry breaking has been studied both by
global (i.e.~averaged over all directions orthogonal to the 
compactified direction) and local quantities. We have shown that
center symmetry in the deformed theory can be completely restored 
in a way which is sometimes qualitatively different from that
of the standard confined phase, as evinced from the expectation value
of local quantities directly coupled to the deformations,
and sometimes in contrast with expectations from the 1-loop effective
potential, since the restoration takes place through long range 
disorder.

Despite this variety of possible restorations, 
our numerical results show that the $\theta$-dependence
of the deformed theory matches, within statistical errors, 
that of the standard confined 
phase in all cases in which center symmetry is completely restored.
On the contrary, a partial restoration of center symmetry 
leads to a $\theta$-dependence which is different from both that
of the confined phase and that of the deconfined phase, 
interpolating in some way between them.

The failure to reproduce predictions for the $\theta$-dependence 
coming from semiclassical computations
(in particular those equivalent to a sort of DFIGA)
can be ascribed, as for the $SU(3)$ results reported in Ref.~\cite{gufo},
to the fact that our inverse compactifications radius is still not large.
On the other hand, the striking agreement with results from the standard
confined phase confirms and reinforces the evidence, already shown 
for $SU(3)$, for a strict relation between the realization
of center symmetry and other relavant non-perturbative features
of Yang-Mills theories. 

Future studies could extend the present investigation in various 
directions. Considering other relevant non-perturbative properties,
such as the spectrum of glueball masses, is a first non-trivial
goal that should be pursued. The extension to large $SU(N)$ gauge
groups is of course another interesting direction.
\\

\emph{Acknowledgement}
Numerical simulations have been performed at the Scientific Computing Center at
INFN-PISA and on the MARCONI machine at CINECA, based on ISCRA Project
IsB18-TDTDYM.


\begin{thebibliography}{95}

\bibitem{Vicari:2008jw} 
  E.~Vicari and H.~Panagopoulos,
  Phys.\ Rept.\  {\bf 470}, 93 (2009)
  [arXiv:0803.1593 [hep-th]].

\bibitem{Hooft-74}
  G.~'t Hooft, 
  Nucl. Phys. B {\bf 72}, 461 (1974).
\bibitem{Witten-80} 
  E.~Witten,
  Annals Phys.\  {\bf 128}, 363 (1980).
\bibitem{Witten-98}
  E.~Witten, 
  Phys. Rev. Lett. {\bf 81}, 2862 (1998).


\bibitem{Lucini:2001ej} 
  B.~Lucini and M.~Teper,
  JHEP {\bf 0106}, 050 (2001)
  [hep-lat/0103027].
\bibitem{DelDebbio:2002xa} 
  L.~Del Debbio, H.~Panagopoulos and E.~Vicari,
  JHEP {\bf 0208}, 044 (2002)
  [hep-th/0204125].
\bibitem{LTW-05} 
  B.~Lucini, M.~Teper, U.~Wenger,
  Nucl. Phys. B {\bf 715}, 461 (2005) 
  [arXiv:hep-lat/0401028].
\bibitem{Ce:2016awn} 
  M.~C\`e, M.~Garc\`ia Vera, L.~Giusti and S.~Schaefer,
  Phys.\ Lett.\ B {\bf 762}, 232 (2016)
  [arXiv:1607.05939 [hep-lat]].


\bibitem{Witten-79}
  E.~Witten, 
  Nucl. Phys. B {\bf 156}, 269 (1979).
\bibitem{Veneziano-79} 
  G.~Veneziano, 
  Nucl. Phys. B {\bf 159}, 213 (1979).


\bibitem{DElia:2003zne} 
M.~D'Elia,
Nucl.\ Phys.\ B {\bf 661}, 139 (2003)
[hep-lat/0302007].

  \bibitem{Giusti:2007tu} 
  L.~Giusti, S.~Petrarca and B.~Taglienti,
  Phys.\ Rev.\ D {\bf 76}, 094510 (2007)
  [arXiv:0705.2352 [hep-th]].
\bibitem{Panagopoulos:2011rb}  
  H.~Panagopoulos and E.~Vicari,
  JHEP {\bf 1111}, 119 (2011)
  [arXiv:1109.6815 [hep-lat]].
\bibitem{Ce:2015qha} 
M.~C\`e, C.~Consonni, G.~P.~Engel and L.~Giusti,
  Phys.\ Rev.\ D {\bf 92}, no. 7, 074502 (2015)
  [arXiv:1506.06052 [hep-lat]].
\bibitem{Bonati:2015sqt} 
  C.~Bonati, M.~D'Elia and A.~Scapellato,
  Phys.\ Rev.\ D {\bf 93}, 025028 (2016)
  [arXiv:1512.01544 [hep-lat]].
\bibitem{Bonati:2016tvi} 
  C.~Bonati, M.~D'Elia, P.~Rossi and E.~Vicari,
  Phys.\ Rev.\ D {\bf 94}, 085017 (2016)
  [arXiv:1607.06360 [hep-lat]].


\bibitem{Gross:1980br} 
  D.~J.~Gross, R.~D.~Pisarski and L.~G.~Yaffe,
  Rev.\ Mod.\ Phys.\  {\bf 53}, 43 (1981).

\bibitem{CDG-78}
  C.~G.~Callan, R.~Dashen, D.~J.~Gross,
  Phys. Rev. D {\bf 17}, 2717 (1978).



\bibitem{KPT-98}
  D.~Kharzeev, R.~D.~Pisarski, M.~H.~G.~Tytgat,
  Phys. Rev. Lett. {\bf 81}, 512 (1998), arXiv:hep-ph/9804221.
\bibitem{BL-07}
  O.~Bergman, G.~Lifschytz, 
  JHEP {\bf 04}, 043 (2007), arXiv:hep-th/0612289.
\bibitem{PZ-08}
  A.~Parnachev, A.~R.~Zhitnitsky,
  Phys. Rev. D {\bf 78}, 125002 (2008), arXiv:0806.1736.



\bibitem{susc_ft} 
  B.~All\'es, M.~D'Elia, A.~Di Giacomo,
  Nucl. Phys. B {\bf 494}, 281 (1997) [Erratum-ibid. B {\bf 679}, 397 (2004)]
  {arXiv:hep-lat/9605013};
  Phys. Lett. B {\bf 412}, 119 (1997) {arXiv:hep-lat/9706016};
  Phys. Lett. B {\bf 483}, 139 (2000) {arXiv:hep-lat/0004020}.
\bibitem{GHS-02}
  C.~Gattringer, R.~Hoffmann, S.~Schaefer,
  Phys. Lett. B {\bf 535}, 358 (2002) {arXiv:hep-lat/0203013}.
\bibitem{DPV-04} 
  L.~Del Debbio, H.~Panagopoulos, E.~Vicari,
  JHEP  {\bf 0409}, 028 (2004) {arXiv:hep-th/0407068}.
\bibitem{Berkowitz:2015aua} 
  E.~Berkowitz, M.~I.~Buchoff and E.~Rinaldi,
  Phys.\ Rev.\ D {\bf 92}, no. 3, 034507 (2015)
  [arXiv:1505.07455 [hep-ph]].

\bibitem{Borsanyi:2015cka} 
  S.~Borsanyi {\it et al.},
  Phys.\ Lett.\ B {\bf 752}, 175 (2016)
  [arXiv:1508.06917 [hep-lat]].

\bibitem{Bonati:2013tt} 
  C.~Bonati, M.~D'Elia, H.~Panagopoulos and E.~Vicari,
  Phys.\ Rev.\ Lett.\  {\bf 110}, 25, 252003 (2013)
  [arXiv:1301.7640 [hep-lat]].


\bibitem{unsal-2}
  E.~Poppitz, T.~Schaefer, M.~Unsal,
  JHEP  {\bf 03}, 087 (2013), arXiv:hep-th/1212.1238.
\bibitem{DElia:2012pvq} 
  M.~D'Elia and F.~Negro,
  Phys.\ Rev.\ Lett.\  {\bf 109}, 072001 (2012)
  [arXiv:1205.0538 [hep-lat]].
\bibitem{DElia:2013uaf} 
  M.~D'Elia and F.~Negro,
  Phys.\ Rev.\ D {\bf 88}, no. 3, 034503 (2013)
  [arXiv:1306.2919 [hep-lat]].


\bibitem{Rossi:2016uce} 
  P.~Rossi,
  Phys.\ Rev.\ D {\bf 94}, 045013 (2016)
  [arXiv:1606.07252 [hep-th]].
\bibitem{Bonanno:2018xtd} 
  C.~Bonanno, C.~Bonati and M.~D'Elia,
  JHEP {\bf 1901}, 003 (2019)
  [arXiv:1807.11357 [hep-lat]].
%
\bibitem{Berni:2019bch} 
  M.~Berni, C.~Bonanno and M.~D'Elia,
  arXiv:1911.03384 [hep-lat].


\bibitem{DiVecchia:1980yfw} 
  P.~Di Vecchia and G.~Veneziano,
  Nucl.\ Phys.\ B {\bf 171}, 253 (1980).
\bibitem{diCortona:2015ldu} 
  G.~Grilli di Cortona, E.~Hardy, J.~Pardo Vega and G.~Villadoro,
  JHEP {\bf 1601}, 034 (2016)
  [arXiv:1511.02867 [hep-ph]].
\bibitem{Bonati:2015vqz} 
  C.~Bonati, M.~D'Elia, M.~Mariti, G.~Martinelli, M.~Mesiti, F.~Negro, F.~Sanfilippo and G.~Villadoro,
  JHEP {\bf 1603}, 155 (2016)
  [arXiv:1512.06746 [hep-lat]].


\bibitem{Unsal:2008ch} 
  M.~Unsal and L.~G.~Yaffe,
  Phys.\ Rev.\ D {\bf 78}, 065035 (2008)
  [arXiv:0803.0344 [hep-th]].

\bibitem{Myers:2007vc} 
  J.~C.~Myers and M.~C.~Ogilvie,
  Phys.\ Rev.\ D {\bf 77}, 125030 (2008)
  [arXiv:0707.1869 [hep-lat]].


\bibitem{Kovtun:2007py} 
  P.~Kovtun, M.~Unsal and L.~G.~Yaffe,
  JHEP {\bf 0706}, 019 (2007)
  [hep-th/0702021 [HEP-TH]].
\bibitem{Unsal:2007vu} 
  M.~Unsal,
  Phys.\ Rev.\ Lett.\  {\bf 100}, 032005 (2008)
  [arXiv:0708.1772 [hep-th]].
\bibitem{Unsal:2007jx} 
  M.~Unsal,
  Phys.\ Rev.\ D {\bf 80}, 065001 (2009)
  [arXiv:0709.3269 [hep-th]].
\bibitem{Shifman:2008ja} 
  M.~Shifman and M.~Unsal,
  Phys.\ Rev.\ D {\bf 78}, 065004 (2008)
  [arXiv:0802.1232 [hep-th]].
\bibitem{Myers:2009df} 
  J.~C.~Myers and M.~C.~Ogilvie,
  JHEP {\bf 0907}, 095 (2009)
  [arXiv:0903.4638 [hep-th]].
\bibitem{Cossu:2009sq} 
  G.~Cossu and M.~D'Elia,
  JHEP {\bf 0907}, 048 (2009)
  [arXiv:0904.1353 [hep-lat]].
\bibitem{Meisinger:2009ne} 
  P.~N.~Meisinger and M.~C.~Ogilvie,
  Phys.\ Rev.\ D {\bf 81}, 025012 (2010)
  [arXiv:0905.3577 [hep-lat]].
\bibitem{Unsal:2010qh} 
  M.~Unsal and L.~G.~Yaffe,
  JHEP {\bf 1008}, 030 (2010)
  [arXiv:1006.2101 [hep-th]].
\bibitem{Thomas:2012ib} 
  E.~Thomas and A.~R.~Zhitnitsky,
  Phys.\ Rev.\ D {\bf 86}, 065029 (2012)
  [arXiv:1203.6073 [hep-ph]].
\bibitem{Poppitz:2012sw} 
  E.~Poppitz, T.~Schaefer and M.~Unsal,
  JHEP {\bf 1210}, 115 (2012)
  [arXiv:1205.0290 [hep-th]].
\bibitem{Thomas:2012tu} 
  E.~Thomas and A.~R.~Zhitnitsky,
  Phys.\ Rev.\ D {\bf 87}, 085027 (2013)
  [arXiv:1208.2030 [hep-ph]].
\bibitem{Poppitz:2012nz} 
  E.~Poppitz, T.~Schaefer and M.~Unsal,
  JHEP {\bf 1303}, 087 (2013)
  [arXiv:1212.1238 [hep-th]].
\bibitem{Misumi:2014raa} 
  T.~Misumi and T.~Kanazawa,
  JHEP {\bf 1406}, 181 (2014)
  [arXiv:1405.3113 [hep-ph]].
\bibitem{Anber:2014lba} 
  M.~M.~Anber, E.~Poppitz and B.~Teeple,
  JHEP {\bf 1409}, 040 (2014)
  [arXiv:1406.1199 [hep-th]].
\bibitem{Bhoonah:2014gpa} 
  A.~Bhoonah, E.~Thomas and A.~R.~Zhitnitsky,
  Nucl.\ Phys.\ B {\bf 890}, 30 (2014)
  [arXiv:1407.5121 [hep-ph]].
\bibitem{Cherman:2016vpt} 
  A.~Cherman, S.~Sen, M.~L.~Wagman and L.~G.~Yaffe,
  Phys.\ Rev.\ D {\bf 95}, 074512 (2017)
  [arXiv:1612.00403 [hep-lat]].
\bibitem{Sulejmanpasic:2016llc} 
  T.~Sulejmanpasic,
  Phys.\ Rev.\ Lett.\  {\bf 118}, 011601 (2017)
  [arXiv:1610.04009 [hep-th]].
\bibitem{Anber:2017rch} 
  M.~M.~Anber and A.~R.~Zhitnitsky,
  Phys.\ Rev.\ D {\bf 96}, 074022 (2017)
  [arXiv:1708.07520 [hep-th]].

\bibitem{Tanizaki:2019rbk}
  Y.~Tanizaki and M.~Unsal,
  arXiv:1912.01033 [hep-th].

\bibitem{Itou:2018wkm}
  E.~Itou,
  JHEP {\bf 1905} (2019) 093
  [arXiv:1811.05708 [hep-th]].

\bibitem{Bergner:2018unx}
  G.~Bergner, S.~Piemonte and M.~Ünsal,
  JHEP {\bf 1811} (2018) 092
  [arXiv:1806.10894 [hep-lat]].

\bibitem{Thomas:2011ee} 
  E.~Thomas and A.~R.~Zhitnitsky,
  Phys.\ Rev.\ D {\bf 85}, 044039 (2012)
  [arXiv:1109.2608 [hep-th]].

  
  
  
\bibitem{unsal-1}
  M.~Unsal,
  Phys. Rev. D {\bf 86}, 105012 (2012), arXiv:1201.6426.

\bibitem{Aitken:2018mbb} 
  K.~Aitken, A.~Cherman and M.~\"Unsal,
  arXiv:1804.06848 [hep-th].


\bibitem{gufo} 
  C.~Bonati, M.~Cardinali and M.~D'Elia,
  Phys.\ Rev.\ D {\bf 98}, 054508 (2018)
  [arXiv:1807.06558 [hep-lat]].


\bibitem{Lucini:2005vg} 
  B.~Lucini, M.~Teper and U.~Wenger,
  JHEP {\bf 0502}, 033 (2005)
  [hep-lat/0502003].


\bibitem{Wilson:1974sk} 
  K.~G.~Wilson,
  Phys.\ Rev.\ D {\bf 10}, 2445 (1974).

\bibitem{Metropolis:1953am} 
  N.~Metropolis, A.~W.~Rosenbluth, M.~N.~Rosenbluth, A.~H.~Teller and E.~Teller,
  J.\ Chem.\ Phys.\  {\bf 21}, 1087 (1953).
\bibitem{Creutz:1980zw} 
  M.~Creutz,
  Phys.\ Rev.\ D {\bf 21}, 2308 (1980).
\bibitem{Kennedy:1985nu} 
  A.~D.~Kennedy and B.~J.~Pendleton,
  Phys.\ Lett.\  {\bf 156B}, 393 (1985).
\bibitem{Creutz:1987xi} 
  M.~Creutz,
  Phys.\ Rev.\ D {\bf 36}, 515 (1987).
\bibitem{Cabibbo:1982zn} 
  N.~Cabibbo and E.~Marinari,
  Phys.\ Lett.\  {\bf 119B}, 387 (1982).

\bibitem{Berg:1981nw} 
  B.~Berg,
  Phys.\ Lett.\ B {\bf 104}, 475 (1981).
\bibitem{Iwasaki:1983bv} 
  Y.~Iwasaki and T.~Yoshie,
  Phys.\ Lett.\ B {\bf 131}, 159 (1983).
\bibitem{Itoh:1984pr} 
  S.~Itoh, Y.~Iwasaki and T.~Yoshie,
  Phys.\ Lett.\ B {\bf 147}, 141 (1984).
\bibitem{Teper:1985rb} 
  M.~Teper,
  Phys.\ Lett.\ B {\bf 162}, 357 (1985).
\bibitem{Ilgenfritz:1985dz} 
  E.~M.~Ilgenfritz, M.~L.~Laursen, G.~Schierholz, M.~Muller-Preussker and H.~Schiller,
  Nucl.\ Phys.\ B {\bf 268}, 693 (1986).

\bibitem{Bonati:2014tqa} 
  C.~Bonati and M.~D'Elia,
  Phys.\ Rev.\ D {\bf 89}, 105005 (2014)
  [arXiv:1401.2441 [hep-lat]].
\bibitem{Cichy:2014qta} 
  K.~Cichy, A.~Dromard, E.~Garcia-Ramos, K.~Ottnad, C.~Urbach, M.~Wagner, U.~Wenger and F.~Zimmermann,
  PoS LATTICE {\bf 2014}, 075 (2014)
  [arXiv:1411.1205 [hep-lat]].
\bibitem{Namekawa:2015wua} 
  Y.~Namekawa,
  PoS LATTICE {\bf 2014}, 344 (2015)
  [arXiv:1501.06295 [hep-lat]].
\bibitem{Alexandrou:2015yba} 
  C.~Alexandrou, A.~Athenodorou and K.~Jansen,
  Phys.\ Rev.\ D {\bf 92}, 125014 (2015)
  [arXiv:1509.04259 [hep-lat]].
\bibitem{Alexandrou:2017hqw} 
  C.~Alexandrou, A.~Athenodorou, K.~Cichy, A.~Dromard, E.~Garcia-Ramos, K.~Jansen, U.~Wenger and F.~Zimmermann,
  arXiv:1708.00696 [hep-lat].
\bibitem{Berg:2016wfw} 
  B.~A.~Berg and D.~A.~Clarke,
  Phys.\ Rev.\ D {\bf 95}, 094508 (2017)
  [arXiv:1612.07347 [hep-lat]].

\bibitem{DiVecchia:1981qi} 
  P.~Di Vecchia, K.~Fabricius, G.~C.~Rossi and G.~Veneziano,
  Nucl.\ Phys.\ B {\bf 192}, 392 (1981).
\bibitem{DiVecchia:1981hh} 
  P.~Di Vecchia, K.~Fabricius, G.~C.~Rossi and G.~Veneziano,
  Phys.\ Lett.\ B {\bf 108}, 323 (1982).

\bibitem{Campostrini:1988cy} 
  M.~Campostrini, A.~Di Giacomo and H.~Panagopoulos,
  Phys.\ Lett.\ B {\bf 212}, 206 (1988).

\bibitem{Necco:2001xg}
    S.~Necco, R.~Sommer,
    Nucl. Phys. B {\bf 622} (2002)
    [hep-lat/0108008]


\end{thebibliography}
\end{document}